\documentclass[10pt,conference]{IEEEtran}

\IEEEoverridecommandlockouts
\ifCLASSOPTIONcompsoc
  \usepackage[nocompress]{cite}
\else
  \usepackage{cite}
\fi
\usepackage{amsmath,amssymb,amsfonts}
\usepackage{algorithmic}
\usepackage{graphicx}
\usepackage{textcomp}
\usepackage{xcolor}
\usepackage{url}
\usepackage{balance}
\usepackage[subrefformat=parens]{subcaption}
\usepackage[caption=false,font=normalsize,
   labelfont=sf,textfont=sf]{subfig}
\usepackage[nameinlink]{cleveref}
\usepackage{combelow}
\usepackage{multicol}
\usepackage{multirow}
\usepackage{booktabs}
\usepackage{ifthen}
\usepackage{xspace}

\usepackage{doubleblind}

\newboolean{showcomments}
\setboolean{showcomments}{true} 
\ifthenelse{\boolean{showcomments}}{
  \newcommand{\nbc}[3]{
    {\colorbox{#3}{\bfseries\sffamily\scriptsize\textcolor{white}{#1}}}%
    {\textcolor{#3}{\textsf\small$\blacktriangleright$\textit{#2}$\blacktriangleleft$}}}
  \newcommand{\todo}[1]{\nbc{TODO}{#1}{blue}\xspace}
}{
  \newcommand{\nbc}[3]{}
  \renewcommand{\todo}[1]{}
}

\newcommand{\Herzigdataset}{Herzig-heuristics Dataset\xspace}
\newcommand{\HerzigRulesPlus}{Herzig Rules+\xspace}

\newcommand{\etal}{\textit{et~al.}\@\xspace}

\begin{document}
\title{Tangling Pull Requests: Curating a Commit Untangling Dataset from Merged PRs
}

\makeatletter
\if@anonymize
  \author{{\bf Anonymous Author(s)}}
\else
\author{\IEEEauthorblockN{Yuki Ueno}
\IEEEauthorblockA{
    \textit{Institute of Science Tokyo}\\
    Yokohama, Japan \\
    ORCID:0009-0005-3166-9457
}
\and
\IEEEauthorblockN{Profir-Petru P\^ar\cb{t}achi}
\IEEEauthorblockA{
    \textit{Institute of Science Tokyo}\\
    Yokohama, Japan \\
    ORCID:0000-0003-4940-6864
}
\and
\IEEEauthorblockN{Takashi Kobayashi}
\IEEEauthorblockA{
    \textit{Institute of Science Tokyo}\\
    Yokohama, Japan \\
    ORCID:0000-0002-2235-9992
}
}
\fi
\makeatother

\maketitle

\begin{abstract}
Composite commits (CC), in which multiple unrelated changes are bundled into a single commit, are frequent in software development and significantly hinder code comprehension and maintenance.
Although machine learning-based methods have been developed to ``untangle'' such commits into smaller, coherent change sets, these methods require large-scale training data with correct untangling labels.
Preparing such datasets is costly and typically requires expert labelling.

In this study, we propose a scalable and cost-effective method for dataset construction by leveraging commits extracted from open-source repositories' pull requests (PRs).
We empirically validated our dataset and found that when applying our filtering rules, PRs that, when viewed as a single commit, are tangled, yet each individual commit on the feature branch is atomic (ideal PRs), increased from $9.5$\% to $55$\%.
This composite commits dataset is more than $5.7$ times larger than previous heuristic-based datasets.

Using our new dataset, we find that the PR-based dataset differs statistically from previous datasets directly constructed using Herzig's proposed heuristics even after accounting for our proposed rules that may alter CC or STS sizes. 
When constructing datasets using the previous heuristics, they differ statistically along dimensions that impact the confidence voters and are likely to impact learning-based approaches.
We validate the impact on the original Herzig \etal method, which used confidence voters across our dataset.

To show that our approach extends to other languages, we also create a Python dataset which we empirically validate, finding comparable rates for ideal PRs ($56.5$\%).

\end{abstract}


\section{Introduction}
\label{sec:intro}

Modern software development practice recommends that developers use Version Control Systems, e.g., git, to record and manage the revision history of source code.
An ideal commit in software development is one that tackles a single task, i.e. has a single purpose, and is semantically cohesive~\cite{book:BerczukAppleton2002_ideal-commit,conf:TaoKim2015MSR_cc-max29}.
Such a commit should handle adding a single new feature, fixing a single bug, or performing a refactoring rather than mix tasks.
We call such commits ``Single Task Commits'' (\textbf{STCs}).
Previous literature, also refers to such commits as ``atomic''.
Developers are encouraged to maintain STCs by the promise of smoother collaboration, other developers can more easily understand the purpose and scope of content changes~\cite{paper:BarnettEtal2015ICSE_stc-merit,conf:DiasEtal2015SANER_EpiceaUntangler,conf:TaoEtal2012FSE_stc-merit}, easier use of tools such as \texttt{git bisect} and easier reversions when necessary~\cite{conf:HayashiSaekiIWPSE2010_easy-revert}.

The state of development practice diverges from this recommendation.
Developers frequently create commits that mix multiple tasks, e.g., a refactor operation together with implementing a feature enabled by the refactoring.
We will refer to such commits as ``Composite Commits'' (\textbf{CCs}); previous work referred to them as ``tangled''.
Developers often resort to CCs due to a lack of time or a lack of awareness of recommended practice.
Herzig \etal~\cite{conf:HerzigZeller2013MSR, journal:HerzigEtal2016ESE} revealed that up to 15\% of bug-fix commits in five Java projects contained multiple unrelated changes.
Tao and Kim~\cite{conf:TaoKim2015MSR_cc-max29} examined four open-source projects and found that one of them contained 29\% composite changes.
CCs cause developers to spend more time understanding the changelog~\cite{conf:HerzigZeller2013MSR, journal:HerzigEtal2016ESE,conf:TaoKim2015MSR_cc-max29,conf:JiangEtal2021ICSE_BugCC47} and also impact defect prediction models~\cite{conf:HerzigZeller2013MSR, journal:HerzigEtal2016ESE}.

To aid developers and fix version histories, researchers developed automatic untangling methods.
These approaches started from heuristic-based, such as in the seminal work by Herzig \etal~\cite{conf:HerzigZeller2013MSR, journal:HerzigEtal2016ESE} or the later Barnett \etal~\cite{paper:BarnettEtal2015ICSE_stc-merit}, through pure-graph based~\cite{conf:PartachiEtal2020FSE_Flexeme}, to graph learning based~\cite{conf:LiEtal2022FSE_UTANGO,conf:FanEtal2024ASE_HDGNN}, and most recently LLM-based~\cite{zhu2026atomizer:arxiv2026a}.
The accuracy of these untangling algorithms has steadily improved, driven by increasingly powerful modelling techniques.
Still, most work uses artificially tangling data according to the heuristics introduced by Herzig \etal~\cite{conf:HerzigZeller2013MSR, journal:HerzigEtal2016ESE}.
While the heuristics are easy to understand and, at a glance, model a reasonable scenario where CCs can occur, they were not checked against recent in-the-wild CC commits.
Indeed, the rules assume implicitly an Agile, Sprint-based development that may be at odds with the current reality of platforms such as GitHub~\cite{article:decan2022use}.
Further, the newer techniques are increasingly data hungry, which requires scaling up the datasets while ensuring they remain representative of the real-world data distributions.

The seminal Herzig \etal~\cite{conf:HerzigZeller2013MSR,journal:HerzigEtal2016ESE} work relied on expert knowledge to curate a golden dataset.
This dataset became a de facto standard for subsequent studies; however, extending it using the same construction, i.e., using experts, is costly and difficult to scale.
Another approach taken by previous work is to reuse the heuristics derived by Herzig \etal and apply them to newly collected data.
This latter approach is taken by Flexeme~\cite{conf:PartachiEtal2020FSE_Flexeme} and UTANGO~\cite{conf:LiEtal2022FSE_UTANGO} for their C\# and Java datasets, respectively.
This heuristic approach does not scale due to the computational cost involved in sequentially navigating a git history and attempting, and even failing, to \texttt{git cherry-pick}.
This rule-based approach has been scaled up.
In HD-GNN~\cite{conf:FanEtal2024ASE_HDGNN} uses a large dataset of 14,000 commits.
However, a smaller, manually-checked dataset was needed to regain confidence in the data and handle complex cases~\cite{conf:FanEtal2024ASE_HDGNN}.
Both the heuristic construction, and especially the manual curation, are difficult to scale up either due to compute cost or expert effort.

To overcome this bottleneck, we hypothesise that, if suitably curated in an automatable way, pull-requests (PRs) are a source of CCs that have their ground truth STCs available, and the addition of GitHub Actions has caused more projects to adopt a PR-driven development model~\cite{article:decan2022use}.
Specifically, we hypothesise that a successfully merged PR can be treated as a CC, while the individual commits within that PR represent the constituent STCs.
A core insight is the recommended developer practice of addressing a single main task within the scope of a single PR.
As na\"ively collecting PRs is unlikely to yield good data, we first define a set of inclusion-exclusion rules for data collection (\Cref{sec:approach}), which we empirically validate (\Cref{sec:rq2}(RQ2)).
After data collection, we validate that our data is, by an overwhelming margin, proper CC-STC pairs.
Our dataset has an 81\% true CC rate and an 83.7\% true STC rate (\Cref{sec:rq1}(RQ1)).
While not 100\%, we remark that this noise level is within tolerance for modern machine learning techniques.
Indeed, previous work has required manual validation as well to regain confidence~\cite{conf:FanEtal2024ASE_HDGNN}.
While we share this weakness, once our rules are defined, our technique scales as any data crawling technique.

Further, we also compare our data to data constructed using the Herzig \etal heuristics (RQ3).
On a random sample of five projects from our Java PR dataset, by applying previously used rules and accounting for our rules that impact CC and STC sizes, we find that CCs created using the previous rules tend to have less dispersed chunks (in terms of min/max file distance), and smaller package distance.
This suggests that the hunks are more clustered, favouring algorithms that propagate information locally.
Meanwhile, our data has a higher probability of spanning multiple methods and packages.
The under-treatment of more spatially separated changes was previously also addressed by Fan \etal~\cite{conf:FanEtal2024ASE_HDGNN}.
We observe that the state of practice has drifted statistically significantly though by a small margin per Cliff's delta since the Herzig \etal study. 
Indeed, the dataset construction proposed then may have become non-representative of in-the-wild data despite the small change.
We demonstrate that this impacts the performance of the originally proposed confidence voters, suggesting that care should be taken that the data be representative (\Cref{sec:rq4}(RQ4)).
We hypothesise that the shift to PR-driven development~\cite{article:decan2022use} may break previous Agile-informed assumptions that have informed previous heuristics: specifically the two-week single author heuristic used by previous work~\cite{conf:PartachiEtal2020FSE_Flexeme,conf:LiEtal2022FSE_UTANGO}.

In summary, the contributions of this paper are as follows:
\begin{itemize}
    \item We propose a novel and automated method for curating a large-scale dataset for Commit Untangling Tasks by leveraging modern development practice regarding pull requests.
    \item We demonstrate that data drift has occurred since the Herzig \etal study, exacerbating the external threat to validity of work that reuses the proposed dataset construction heuristics.
    \item While the drift we observe is small, it impacts a locality assumption. We show how this risk is realised in the original confidence voters.
\end{itemize}
\section{Related Work}\label{sec:background}

In this section, we provide an overview of previous work tackling the automatic commit untangling problem.
Early work focused on heuristic approaches employing static analysis, which later evolved to presenting static analysis results as graph structures and either employing graph algorithms directly or, most recently, graph learning algorithms.
We then present the evolution of datasets enabling CCs decomposition work as well as existing research that considers PRs as a source of data.

\textbf{Heuristic and Feature-based Approaches.} The foundational work by Herzig \etal~\cite{conf:HerzigZeller2013MSR, journal:HerzigEtal2016ESE} first systematically documented and addressed the problem of CCs.
Their focus was on attesting the existence of CCs and their impact on bug detection algorithms through the bias they introduce.
They are also the first to propose an automatic commit untangling approach.
This approach combines five ``Confidence Voters'' (e.g., File Distance, Change Couplings) to estimate the relationship between code changes based on heuristics.
These voters effectively defined a similarity score that enabled clustering changes.
Herzig \etal's work established the prevalence of CCs and their negative impact, paving the way for subsequent research.
Follow-on early work continued to require manually defined features and heuristics to cluster changes \cite{conf:DiasEtal2015SANER_EpiceaUntangler}.

\textbf{Graph-based Approaches on Local Context.} Shifting away from heuristic based approaches, the next line of research adapted existing structured views of source-code to the multi-version setting.
Initial work here focused on naïve applications of graph clustering to the multi-version representation (Flexeme~\cite{conf:PartachiEtal2020FSE_Flexeme}).
In later work, Li \etal~\cite{conf:LiEtal2022FSE_UTANGO} employ and adapt the multi-version graph representation to enable GCN-based learning approaches further improving untangling performance.
To enable scalability, the multi-version representation was often sliced to a core changeset, obscuring longer-range, global dependencies that could inform untangling algorithms, especially machine learning based approaches.

To address the above issue, the current state-of-the-art (HD-GNN~\cite{conf:FanEtal2024ASE_HDGNN}) augments the multi-version graph with hierarchical edge information at various levels of abstraction (classes, methods, variables).
To better use this new data, HD-GNN employs a heterogeneous GNN that is aware of the different levels-of granularity.
This enables HD-GNN to use previously ``hidden dependencies'' and handle more difficult untangling cases.

In parallel to the HD-GNN direction, Xu \etal~\cite{jarnal:XuEtal2025_attributed-graph} consider a two-phase approach.
First they want to avoid over-segmentation.
They classify if a commit is likely to be tangled.
If so, they cluster using affinity propagation assuming that a single line of change is atomic.
They recognised the previously ignored textual semantics, and expose this to their model by considering an attributed graph and by maintaining source-code line information to use through a vector embedding.

Crucially for our work, these advanced, data-hungry models require increasingly larger and more diverse datasets, which we aim to provide.

\textbf{LLM-based Approaches.} Most recently, Zhu \etal~\cite{zhu2026atomizer:arxiv2026a} build upon the previous graph approaches by exposing an initial clustering to an agentic set-up.
These are iteratively refined until the agents are confident in the clustering.
The final clusters are offered as an untangle recommendation.
This showed marked improvement especially in cases that span many nodes.
This translates to improved performance for large change-sets.

While the LLM approach does not have a training phase, evaluation on datasets that have drifted from the state of practice may obscure its expected real-world performance.

\textbf{Datasets for Commit Untangling.} The most influential dataset in commit untangling research was created by Herzig \etal~\cite{conf:HerzigZeller2013MSR, journal:HerzigEtal2016ESE}.
They constructed this dataset by manually inspecting thousands of commits from five open-source Java projects to classify them as STCs or CCs.
They also defined a set of heuristics, whereby starting from a set of STCs, a dataset of CCs can be artificially constructed to mimic expected development scenarios.
Due to its careful construction and public availability, this dataset has been reused as a benchmark in numerous subsequent studies \cite{conf:MuylaertDe2018SCAM_PdgSlicing,conf:LiKobayashi2021SS_TBCNN,conf:LiEtal2022FSE_UTANGO}.

The creation of this dataset, while foundational, also illustrated the challenges inherent in manual curation.
The process was resource-intensive and highlighted the subjective nature of classifying some change sets, even for experts.
Furthermore, as this dataset was created in 2013, it may not fully represent the practices and languages used in contemporary software development.
Subsequent work has employed the methodology introduced by Herzig \etal\ either directly to create new datasets~\cite{conf:ShenEtal2021FSE_SmartCommit,conf:PartachiEtal2020FSE_Flexeme ,conf:LiEtal2022FSE_UTANGO}, or by reusing existing public datasets that have been created using this technique~\cite{jarnal:XuEtal2025_attributed-graph}.
That manual work required researchers to create seed STCs and later validate the quality of the generated data.
This underlines the importance of developing more scalable methods for dataset creation.
Further, the methodology relies on the state of practice not drifting from the time of the original seminal work~\cite{conf:HerzigZeller2013MSR, journal:HerzigEtal2016ESE}, which may not, necessarily, be a reasonable assumption.
Our research directly tackles this dataset creation bottleneck by aiming to be scalable and mining directly from in-the-wild data both STCs and CCs.

\textbf{Pull Requests as a Source for Labeled Data.} In modern, collaborative software development, the pull request (PR) based workflow is a standard practice on platforms like GitHub and GitLab.
A PR is a mechanism for a developer to propose changes to a repository.
Other team members typically review these changes before being integrated (merged) into the main codebase.
Crucially, a PR is generally created to address a single conceptual task, such as fixing a specific bug, adding a new feature, or performing a significant refactoring \cite{conf:ZhuEtal2016_pr-system} while the review process may induce further changes to support that conceptual task.

This common development practice provides a natural, albeit noisy, source of labelled data for commit untangling.
The overall change encompassed by a merged PR corresponds well to the concept of a Composite Commit, a set of changes aimed at a single, high-level goal with side-tasks required to enable the change.
Concurrently, the sequence of smaller, incremental commits made by the developer within that PR often corresponds to STCs, which are the individual steps taken to achieve the larger goal.
For instance, a developer might first refactor a class in one commit, then add a new method in a second commit, and finally fix a minor bug in a third, all within the same PR to implement a new feature.

Of course, not all PRs fit this idealized model.
Some PRs may be too large, or their internal commits may not be cleanly separated.
This observation motivates our approach in \Cref{sec:approach}, where we introduce a set of filtering conditions designed to identify PRs that closely adhere to the ``one PR, one task'' principle, thereby allowing for the curation of a high-quality dataset.
Later, \Cref{sec:rq1}(RQ1) and \Cref{sec:rq2}(RQ2) validate how well our filtering conditions perform at generating CC and STC data.
By leveraging the inherent structure of the PR workflow, we can move beyond the limitations of manual dataset creation.
\section{Approach}\label{sec:approach}

Software development on platforms such as GitHub\footnote{https://github.com/} and GitLab\footnote{https://about.gitlab.com/} tends to be Pull Request (PR) based~\cite{conf:OrtuEtal2019ACM_pr-merge-emotion}.
Developers can contribute to a project by submitting changes to the source code through PRs~\cite{conf:OrtuEtal2019ACM_pr-merge-emotion}.
Once created, PRs are reviewed by other developers and then closed if the changes are inappropriate or merged if they are appropriate.
A merged PR is considered ``successful'' because it is a valid code change~\cite{conf:RahmanRoy2014pr-success} typically associated with a single high-level task.

Our approach automatically curates a dataset for commit untangling by processing merged PRs from GitHub.
We begin by collecting a large pool of merged PRs that meet our repository-level requirements.
Then, we apply a series of eight heuristic-derived filtering conditions to this pool.
The PRs that pass all filters are considered sufficiently high-quality CC candidates, and their constituent commits are treated as the corresponding STCs, forming the final ``(tangled, atomic)'' data pairs.
We empirically validate this assumption in \Cref{sec:rq1} and \Cref{sec:rq2}.

We only take PR-Commits that are disjoint changes and whose sum constitutes a merge commit.
This simplifies the construction of ground truth data as we do not need to consider overlapping changes and partial visibility.
We consider the following four requirements for data collection:

\vspace{2mm}
\noindent
\textbf{Req. 1: Are public and use permissive licenses.}\,
\hangindent=5mm
We focus our data collection on open-source projects that allow the reuse of their code exposed through PR data, such as the Apache License 2.0.

\noindent
\textbf{Req. 2: Target only a specific language.}\,
\hangindent=5mm
We want to facilitate comparison with previous data construction and untangling using the Herzig \etal heuristics~\cite{conf:HerzigZeller2013MSR,journal:HerzigEtal2016ESE}.
The target languages in this paper are Java and Python.
Since both languages are popular on GitHub, it is possible to collect a large amount of PR data.
Java is the most common programming language targeted in datasets in existing studies~\cite{conf:HerzigZeller2013MSR,journal:HerzigEtal2016ESE,conf:LiKobayashi2021SS_TBCNN,conf:ShenEtal2021FSE_SmartCommit,conf:LiEtal2022FSE_UTANGO,conf:FanEtal2024ASE_HDGNN}.

\noindent
\textbf{Req. 3: Have a minimum star limit.}\,
\hangindent=5mm
To ensure sufficient activity for data collection, and using results from previous work that investigated the relationship between open-source projects and the number of stars on GitHub~\cite{article:HudsonMarco2018_pr-star}, we filter projects by the number of stars.
We set the thresholds at 3,000 stars for Java projects and 6,000 stars for Python projects.
The higher star requirement for Python is due to a larger pool of available projects as the language is more popular.

\noindent
\textbf{Req. 4: Are merged before $\textrm{2024-05-15 23:59 UTC}$.}\,
\hangindent=5mm
While the methodology does not need a strict cut-off and the dataset can become a ``living'' dataset, to facilitate reproduction, we set a merge cut-off of \textrm{2024-05-15 23:59 UTC}.
Future work can also consider cut-off dates as a means to include/exclude AI-generated PRs.

\vspace{2mm}

Thus far, we detailed how to select repositories and which PRs to consider.
We next consider the criteria that apply to PRs and individual commits in a PR.
Developers branch off from a development tree before requesting that their work tree be merged back to the remote tree in a PR.
This local branch can have multiple commits.
We will refer to the individual commits on a PR branch as \textbf{PR-Commits}.
Of the collected merged PRs and PR-Commits in each of the PRs, we included those that meet all of the eight conditions that we now describe.

As we described, all the eight conditions are heuristic-based.
Most previous CC creation rules are based on heuristics~\cite{conf:HerzigZeller2013MSR,journal:HerzigEtal2016ESE,conf:PartachiEtal2020FSE_Flexeme,conf:LiEtal2022FSE_UTANGO}, such as file co-change coupling, and rigorous syntactic and semantic analysis incurs enormous time costs.

We have eight filtering conditions, which can be broadly categorized into four rule categories:

\vspace{2mm}
\noindent
\textbf{Category 1: Approach Scalability.}
\hangindent=5mm

For increasingly data-hungry methods, scalable dataset collection is crucial.
At the same time, we wish to remain representative of real data.
To reduce computational costs, the size of the PR should not be too large.
Also, the larger the change size in a PR, the more likely it is that code smells will be present~\cite{article:AzeemEtal2024_pr-codesmell}, and PR-based projects tend to accept smaller PRs~\cite{article:decan2022use}.
This suggests such rules are unlikely to bias our data.
There are three conditions in this category:

\begin{itemize}
    \item \textbf{Cond. 1}: Contain between two and five PR-Commits.
    \item \textbf{Cond. 2}: Have 100 or fewer LOC changed.
    \item \textbf{Cond. 3}: Have three files or fewer changed.
\end{itemize}

Regarding condition 1, a CC must consist of more than one task, thus our lower bound is two by definition.

\noindent
\textbf{Category 2: Simplify Ground Truth Construction.}

If changes overlap between STCs, we need manual intervention to label the data.
This is due to the ambiguity of the tasks being higher in such circumstances.
Further, this may indicate refinement or fixing of previous patches rather than new tasks, further complicating the ground truth.
Therefore, we adapt that intuition into the following condition:

\begin{itemize}
    \item \textbf{Cond. 4}: Do not contain overlapping source code changes between PR-Commits.
\end{itemize}

\noindent
\textbf{Category 3: Data quality.}

Software developers make commits with commit messages in natural languages to summarise the change contents and the change intents\cite{conf:MockusVotta2000ICSME_CMisEssential,conf:BuseWeimer2010ASE_CMandDocument,conf:LiuEtal2018ASE_NMT}.
The worse the quality of commit messages, such as too generic, e.g., ``fixed bug'', or duplicated, the worse the quality of commits\cite{conf:AgrawalEtal2015SE4HPCS_GenericCM,conf:ChenGoldin2020ICOIAC_DuplicateCM}.
While we need to ensure the quality of a PR-Commit as an STC, a thorough evaluation may be prohibitive at scale.
We opt for the following simple to check conditions:

\begin{itemize}
    \item \textbf{Cond. 5}: Contain no duplicate commit messages.
    \item \textbf{Cond. 6}: Contain no empty commit messages.
    \item \textbf{Cond. 7}: Contains no review comments between PR-Commits.
\end{itemize}

(Cond. 5) If multiple PR-Commits share the same message, it becomes difficult to ascertain the unique purpose of each commit.
This situation could indicate improperly split commits or a lack of distinct, incremental steps.
To ensure each PR-Commit represents a clear, singular task, we exclude such cases.

(Cond. 6) A commit message is the primary record of a developer's intent for a change.
An empty message suggests that the change may lack a clear, singular purpose or is a trivial change not classifiable as a meaningful STC.
We filter these out to maintain the semantic quality of our STC candidates.
Both of the above conditions signal potentially low quality data.

(Cond. 7) Review comments that appear chronologically between the commits of a PR often indicate a dependency.
For example, a later commit may be created specifically to address feedback on an earlier one.
Such a sequence does not represent a set of independent STCs, but rather an iterative refinement process.
To ensure that the constituent commits are as independent as possible, we exclude PRs that contain this type of interactive, review-driven development.

\noindent
\textbf{Category 4: Single Authorship.}
\hangindent=5mm

Herzig \etal note that one common pattern for tangled changes involves consecutive commits from the same author\cite{conf:HerzigZeller2013MSR,journal:HerzigEtal2016ESE}.
We adopt this principle by focusing on single-author PRs.
This filter simplifies the development context, as a sequence of commits from one person is more likely to represent a continuous thought process for a single task.
PRs with multiple authors can involve complex hand-offs that do not fit our idealized model of a CC composed of independent STCs.
CCs are essentially created unintentionally by a single developer.

\begin{itemize}
    \item \textbf{Cond. 8}: Are created by a single author.
\end{itemize}

\section{Evaluation}\label{sec:eval}

In this section, we first validate the quality of our dataset by manually checking the true rate of CCs and STCs (RQ1).
We then validate the impact of our filtering rules by exploring the rate of CCs and STSs among PRs that were eliminated from our dataset (RQ2).
Next, we assess if there is statistical difference between datasets constructed using our proposed method and previous approaches (RQ3).
Finally, we check if the data drift we observe in RQ3 impacts the performance of a confidence voters based approach by replicating the Herzig \etal commit untangling approach (RQ4).
Therefore, we seek to answer the following research questions:
\begin{itemize}
    \item \textbf{RQ1}: After our filtering, are individual commits on a PR branch STC? Do these PRs form CCs when considering the full patch?
    \item \textbf{RQ2}: Does following our filtering rules increase the quality of the dataset in terms of ideal PRs: the full PR is CC and the individual commits are STC?
    \item \textbf{RQ3}: Are datasets constructed using our approach and previous heuristics statistically different along structural dimensions: number of hunks, LOC, file distance, and package distance?
    \item \textbf{RQ4}: What is the performance of Herzig \etal's confidence voters on both our proposed PR dataset and the \Herzigdataset?
\end{itemize}

\subsection{{\sf RQ1}: Are PRs a good source of (CC, STC) paired data?}
\label{sec:rq1}

The core assumption of our automated curation method is that PRs after the filtering described in \Cref{sec:approach} can serve as valid CCs, with their constituent commits acting as STCs.
To validate this key assumption, we conducted a manual labelling study.
This RQ addresses the following questions: \textit{After our filtering, are individual commits on a PR branch STC? Do these PRs form CCs when considering the full patch?}

\subsubsection{Experimental Methodology}

We first collected a list of merged PRs from Java repositories that satisfied the repository-level requirements (e.g., public, high star count) detailed in \Cref{sec:approach}.
We used the PyGitHub \footnote{https://github.com/PyGithub/PyGithub} library to interact with the GitHub REST API \footnote{https://docs.github.com/en/rest?apiVersion=2022-11-28} in order to collect the data.
Subsequently, we applied our eight filtering conditions to this initial pool of PRs to distill it into our final dataset.

\Cref{table:PR_data_adopted_java} summarizes the results of the data collection and our filtering process.
Our method curated 15,320 PRs, which we treat as CCs, from an initial pool of 893,131 merged PRs across 1,092 repositories.
\begin{table}[t]
    \caption{Repository and PR data before and after filtering (Java).}
    \label{table:PR_data_adopted_java}
    \centering
    \normalsize{
        \begin{tabular}{ccc}
            \hline
             & Repositories & Merged PRs \\
            \hline
            Before filtering & 1,092 & 893,131 \\
            After filtering & 606 & 15,320 \\
            \hline
            Ratio & 1 of 1.8 & 1 of 58.3 \\
            \hline
        \end{tabular}
    }
\end{table}

\textbf{Sampling for manual validation:} For our manual analysis, we sampled from the dataset of 15,320 curated PRs described in \Cref{table:PR_data_adopted_java}.
To prepare this dataset for all research questions, we applied three additional filtering conditions, resulting in a refined pool of 12,571 PRs.
The additional filters were needed to ensure code compiles for static analysis as well as ensure that the tasks that are tangled both relate to source files.
The conditions were as follows:
\begin{itemize}
    \item The PR must contain two or more \textbf{chunks} in \texttt{.java} files.
    This is a prerequisite for our untangling experiment, which operates on Java file chunks and requires at least two units to perform a split.
    A chunk is a contiguous block of added, deleted lines of code within a file.
    \item There must be no overlapping chunks between the constituent PR-Commits.
    This condition was necessary due to a limitation in our metadata extraction script, which could not correctly assign ownership of overlapping chunks to a unique PR-Commit.
    \item The number of chunks across all PR-Commits and the number of chunks in the merged PR commits are identical.
    This ensures the validity of our evaluation metric (precision as defined in Herzig \etal~\cite{conf:HerzigZeller2013MSR, journal:HerzigEtal2016ESE}), which requires a clear mapping of chunks and would not function correctly if the total number of units changes.
\end{itemize}

The distribution of PRs in this refined pool of 12,571 candidates is shown in \Cref{table:refined_dist_adopted}.
From this pool, we randomly sampled 100 PRs for manual labelling.
We chose this number to ensure statistical reliability.
This sample size is determined by Cochran's formula~\cite{cochran1934distribution} for estimating a population proportion.
The required sample size is 96 for a 95\% confidence level and a 10\% margin of error.
The previous study~\cite{conf:PartachiEtal2020FSE_Flexeme} manually verified 30 STC samples.
They identified a 10\% error rate where samples were not true STCs.
We adopted the same 10\% margin of error as a permissible threshold.

The sample consisted of 85 PRs with two PR-Commits, 10 with three, 4 with four, and 1 with five, for a total of 221 PR-Commits to be evaluated.

\begin{table}[t]
    \caption{Distribution of adopted PRs in the Refined Pool and Sample.}
    \label{table:refined_dist_adopted}
    \centering
    \normalsize{
        \begin{tabular}{lrrrrr}
            \hline
            & \multicolumn{5}{c}{\# PR-Commits} \\ \cline{2-6}
            Dataset & 2 & 3 & 4 & 5 & Total \\
            \hline
            Refined & 10,526 & 1,577 & 359 & 109 & 12,571 \\
            Sampled & 85 & 10 & 4 & 1 & 100 \\
            \hline
        \end{tabular}
    }
\end{table}

\textbf{Labelling Process:}
The two authors independently labelled the 100 sampled PRs and their 221 constituent PR-Commits.
For each item, they answered a binary (True/False) question:
\begin{itemize}
    \item \textbf{For each PR (CC candidate)}: Does this PR consist multiple independent tasks (e.g., fixing one bug and one refactoring)?
    \item \textbf{For each PR-Commit (STC candidate)}: Does this commit represent a single, atomic development step?
\end{itemize}
We classify a change as ``CC'' if it involves multiple tasks, such as testing and implementation, documentation and implementation, or refactoring with different intentions.
Renaming the same method across various call sites constitutes a single task while renaming multiple methods represents a task for each method.
Commits that cannot be classified as either CC or STC due to various reasons, such as no visible changes, are labelled as ``Unknown''.
This process resulted in a total of 321 labels (100 for PRs, 221 for PR-Commits) from each annotator.
To account for order effects, the labelling order differed between the two annotators.

\textbf{Inter-Rater Reliability:}
To measure the consistency of the labelling, we calculated the Cohen's kappa coefficient ($\kappa$)~\cite{article:Tarald1991_cohen-kappa} on the 321 pairs of labels.
The resulting kappa value was \textbf{0.76}, which indicates a ``substantial'' level of agreement between the two annotators.
The raw agreement rate was also high, with the annotators assigning the same label in 299 out of 321 cases (93.1\%).

\textbf{Labelling Outcome:}
\Cref{table:labelling_results_java_adopted} summarises the results of the manual labelling from both annotators.
We define the ``True CC Rate'' as the proportion of PRs labelled as True, and the ``True STC Rate'' as the proportion of PR-Commits labelled as True.
The results show that human experts deemed a high percentage of the automatically curated data points as valid.

\begin{table}[t]
    \caption{Manual Labelling Results (Java).}
    \label{table:labelling_results_java_adopted}
    \centering
    \normalsize{
        \begin{tabular}{llcc}
            \hline
            Annotator & Metric & Value & Percentage \\
            \hline
            \multirow{3}{*}{1st Author} & True CC Rate & 86 / 100 & 86.0\% \\
                                    & True STC Rate & 180 / 221 & 81.4\% \\
                                    & Ideal PR Rate & 57 / 100 & 57.0\% \\
            \hline
            \multirow{3}{*}{2nd Author} & True CC Rate & 76 / 100 & 76.0\% \\
                                    & True STC Rate & 190 / 221 & 86.0\% \\
                                    & Ideal PR Rate & 53 / 100 & 53.0\% \\
            \hline
        \end{tabular}
    }
\end{table}

\begin{table}[t]
    \caption{Repository and PR data before and after filtering (Python).}
    \label{table:PR_data_adopted_python}
    \centering
    \normalsize{
        \begin{tabular}{ccc}
            \hline
             & Repositories & Merged PRs \\
            \hline
            Before filtering & 1,262 & 1,353,615 \\
            After filtering & 863 & 40,303 \\
            \hline
            Ratio & 1 of 1.5 & 1 of 33.6 \\
            \hline
        \end{tabular}
    }
\end{table}

Our manual validation, supported by a substantial inter-rater agreement ($\kappa=0.76$), demonstrates that our automated curation method produces sufficiently high quality datasets.
The sampled PRs were identified as valid Composite Commits in 76.0 - 86.0\% (81.0\%) of cases, and their constituent commits were confirmed as valid Single Task Commits in 81.0 - 86.0\% (83.7\%) of cases.
This result provides strong evidence that our core assumption is sound.

While some noise in the result is present in the automatically curated data, it is within the tolerance limits of modern machine learning techniques. 
Local density techniques can adjust a radius based threshold, while machine learning or spectral methods can demote the edge weight.
Previous work has shown upwards of $70$\% accuracy on GNN tasks without noise treatment at comparable noise levels while also proposing a noise tolerant method~\cite{article:dai2021nrgnn}.
This confirms the dataset's high quality for practical applications.

Additionally, we have applied the same protocol to Python.
\Cref{table:PR_data_adopted_python} summarised the collected data before and after filtering, while \Cref{table:labelling_results_python_adopted} shows the results of applying the same manual labelling protocol.
The resulting Cohen's kappa is $0.62$, indicating substantial and the exact agreement was $294/325 (90.5\%)$.
This shows that our approach can extend beyond Java.

\begin{table}[t]
    \caption{Manual Labelling Results (Python).}
    \label{table:labelling_results_python_adopted}
    \centering
    \normalsize{
        \begin{tabular}{llcc}
            \hline
            Annotator & Metric & Value & Percentage \\
            \hline
            \multirow{3}{*}{1st Author} & True CC Rate & 88 / 100 & 88.0\% \\
                                    & True STC Rate & 190 / 221 & 84.4\% \\
                                    & Ideal PR Rate & 58 / 100 & 58.0\% \\
            \hline
            \multirow{3}{*}{2nd Author} & True CC Rate & 79 / 100 & 79.0\% \\
                                    & True STC Rate & 197 / 221 & 87.6\% \\
                                    & Ideal PR Rate & 55 / 100 & 55.0\% \\
            \hline
        \end{tabular}
    }
\end{table}

\subsection{{\sf RQ2}: Do our filtering rules increase the rate of ideal PRs?}
\label{sec:rq2}

\begin{table}[t]
    \caption{Distribution of eliminated PRs in the Refined Pool and Sample.}
    \label{table:refined_dist_eliminated}
    \centering
    \normalsize{
        \begin{tabular}{lrrrrrr}
            \hline
            & \multicolumn{6}{c}{\# PR-Commits} \\ \cline{2-7}
            Dataset & 2 & 3 & 4 & 5 & others & Total \\
            \hline
            Refined & 109k & 58k & 34k & 23k & 632k & 856k\\
            Sampled & 42 & 30 & 12 & 16 & 0 & 100 \\
            \hline
        \vspace{-1em}
        \end{tabular}
    }
\end{table}

To empirically validate the impact of our filtering rules, we reused the manual protocol from RQ1; however, we retain only the following filtering rules: the PR has between two and five PR-Commits.
We want to avoid degenerate cases of a single commit or extreme numbers of commits.
It changes at least one file as otherwise there is no data to assess, and there is no overlapping change among PR-Commits.
We then compare these results against those observed in RQ1.

On our sample of 100 PRs and 302 PR-Commits, we found that, while the CC and STC rates were lower: 61\% vs 81\%, and respectively 72\% vs 84\%, the main impact on ideal PR rate which falls from 55\% to 9.5\%.
The resulting kappa value was \textbf{0.69}, which also indicates a ``substantial'' level of agreement between the two annotators.
Additionally, as \Cref{table:refined_dist_eliminated} shows, while we examined the PRs with less than six PR-Commits and highest chance to pass our quality check, data quality still significantly dropped.

\subsection{{\sf RQ3}: Comparative Analysis of Dataset Construction Rules}
\label{sec:rq3}

The seminal work by Herzig \etal\cite{conf:HerzigZeller2013MSR, journal:HerzigEtal2016ESE} provided the community with a foundational benchmark for commit untangling.
They also provided three STC combination rules which are used to create synthetic CCs.
P\^{a}r\cb{t}achi \etal~\cite{conf:PartachiEtal2020FSE_Flexeme} augmented Herzig \etal's rules to artificially create CCs from commits in open-source projects.
Li \etal\cite{conf:LiEtal2022FSE_UTANGO} reused the augmented rules to construct a new dataset.
We randomly select repositories from our PR dataset, and then make two datasets: one using our PRs (Ours) and one using augmented Herzig rules (\HerzigRulesPlus).
Additionally to the previously used rules, we add that CCs should contain between two and five STCs, must not change more than 100 LOC, and must not change more than three files.
Without these rules we may inadvertently bias our LOC and number of Chunks metric.

To understand how our automatically curated dataset compared to Herzig \etal's approach, we conducted a comparative statistical analysis.
This RQ addresses the question: \textit{Is our dataset statistically similar to a dataset created using Herzig \etal's method?}

\subsubsection{Experimental Methodology} \ 

\noindent\textit{\textbf{Baseline:}}
We compare our rules against the \HerzigRulesPlus, which we augment from the original rules to remove bias that may be from our filtering rules.
They considered sequences of commits such that, starting from the earliest one in the sequence, each following ones:
(1) Are created by the same developer within 14 days;
(2) Have at least one pair of changed files with a substantial prefix match in their paths;
(3) Contain files that are frequently co-changed.
We also adopt another rule used by P\^{a}r\cb{t}achi \etal~\cite{conf:PartachiEtal2020FSE_Flexeme}:
(4) Contain only one of the following keywords in the commit message (`FIX', `FIXES', `FIXED', `IMPLEMENTS', `IMPLEMENTED', `IMPLEMENT', `BUG', `FEATURE').
\HerzigRulesPlus incorporates Cond. 1, 2, and 3 from our rules to remove their bias.
Both our conditions and previously used rules set the minimum number of chunks and STCs as two.
The maximum number of STCs five is also consistent with the setting by P\^{a}r\cb{t}achi \etal

For this RQ, we consider six metrics:
the number of chunks per STC (or PR-Commit in our case);
the number of changed LOC per STC;
the minimum file distance per CC;
the maximum file distance per CC;
the minimum package distance per CC;
the maximum package distance per CC.
File distance and package distance were selected from the five ``Confidence Voters'' detailed in the original work by Herzig \etal\cite{conf:HerzigZeller2013MSR, journal:HerzigEtal2016ESE}, which we also employ in \Cref{sec:rq4}.
We do so as to anticipate potential impact on previous work.
Regarding the number of changed files, the metric is calculated per CC, as there are cases where the same file is changed across different STCs.

\noindent\textit{\textbf{Procedure:}}
To ensure a sufficient sample size for statistical testing, we focused exclusively on Java repositories that have at least 100 CCs, as determined by our rules.
From this set, we randomly selected five repositories.
Our rules curated 598 CCs and 1,295 STCs, and \HerzigRulesPlus: 236 CCs and 534 STCs.

We compared the data from our rules against that curated by \HerzigRulesPlus across three key structural dimensions.
First, we consider size related metrics such as the number of chunks and changed lines of code (LOC) per STC.
Second, we investigate the spatial locality of CCs by analysing the distributions of four scores: (min/max) file distance and (min/max) package distance per CC.

To assess the differences between the distributions for each metric, we employed the Kolmogorov-Smirnov (KS) test at a significance level of 0.05.
This non-parametric test is suitable for determining whether two independent samples are drawn from the same distribution.

\subsubsection{Results}
\Cref{table:dataset_comparison} presents a summary of the descriptive statistics for the core metrics across both rules.
At a glance, the rules exhibit notable differences in their central tendency and standard deviation.

\begin{table}[t]
    \caption{Statistical Data Comparison by the Rules. FD: File Distance, PD: Package Distance. n: negligible, s: small.}
    \label{table:dataset_comparison}
    \centering
    \normalsize{
    \begin{tabular}{llrrr}
        \toprule
        \textbf{Metric} & \textbf{Rules} & \textbf{Ave.} & \textbf{Med.} & \textbf{Cliff's $\delta$} \\
        \midrule
        \multirow{2}{*}{\# Chunks} & Ours & 4.34 & 3.00 & \multirow{2}{*}{-0.17 (s)} \\
                                        & Herzig's + & 5.53 & 5.00 & \\
        \midrule
        \multirow{2}{*}{Chg. LOC} & Ours & 14.46 & 11.00 & \multirow{2}{*}{-0.13 (n)} \\
                                        & Herzig's + & 16.88 & 15.00 &  \\
        \midrule
        \multirow{2}{*}{min FD} & Ours & 27.18 & 3.00 & \multirow{2}{*}{0.10 (n)} \\
                                        & Herzig's + & 11.10 & 2.50 &  \\
        \midrule
        \multirow{2}{*}{max FD} & Ours & 274.60 & 109.00 & \multirow{2}{*}{0.17 (s)} \\
                                        & Herzig's + & 124.24 & 69.50 &  \\
        \midrule
        \multirow{2}{*}{min PD} & Ours & 4.74 & 3.00 & \multirow{2}{*}{0.21 (s)} \\
                                        & Herzig's + & 2.83 & 1.00 &  \\
        \midrule
        \multirow{2}{*}{max PD} & Ours & 5.73 & 7.00 & \multirow{2}{*}{0.16 (s)} \\
                                        & Herzig's + & 4.32 & 2.00 &  \\
        \bottomrule
    \end{tabular}
    }
\end{table}

\begin{figure*}[t]
    \centering
    \subcaptionbox{File Distance\label{fig:file_dist}}[.19\linewidth]{\includegraphics[width=.50\linewidth]{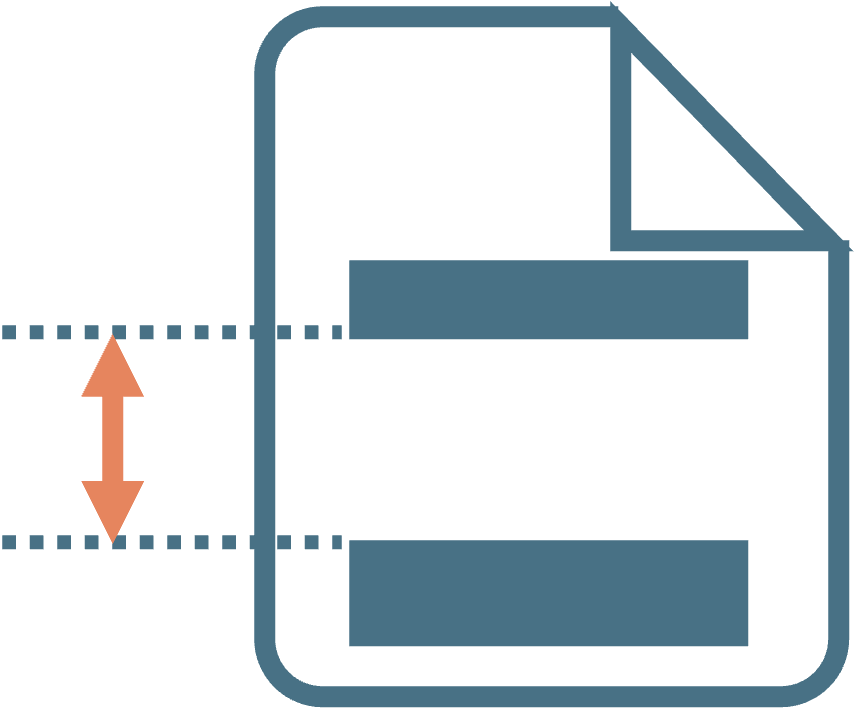}}
    \subcaptionbox{Package Distance\label{fig:pkg_dist}}[.19\linewidth]{\includegraphics[width=.75\linewidth]{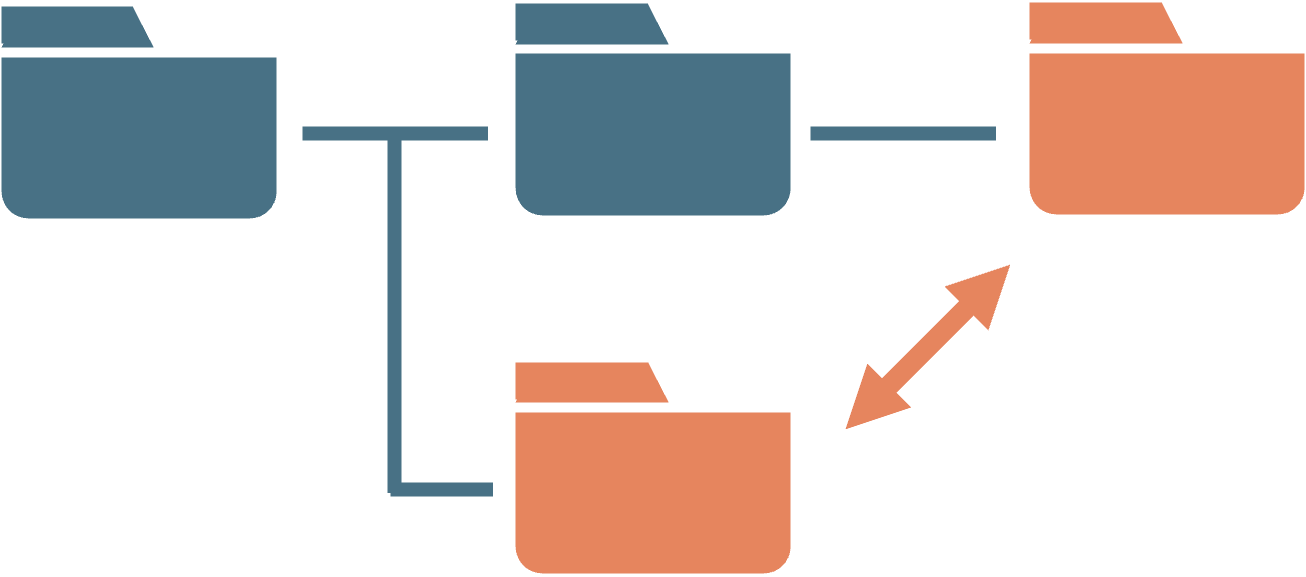}}
    \subcaptionbox{Call Graph\label{fig:call_graph}}[.19\linewidth]{\includegraphics[width=.75\linewidth]{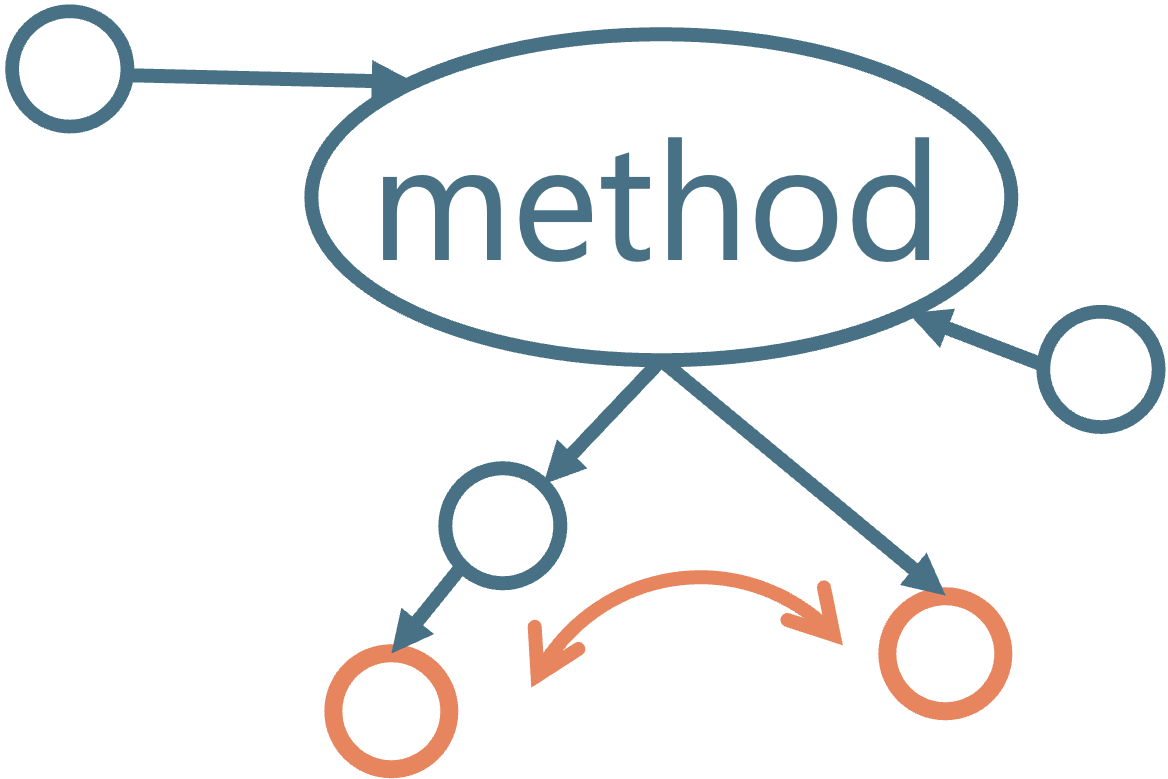}}
    \subcaptionbox{Change Coupling\label{fig:change_coupling}}[.19\linewidth]{\includegraphics[width=.75\linewidth]{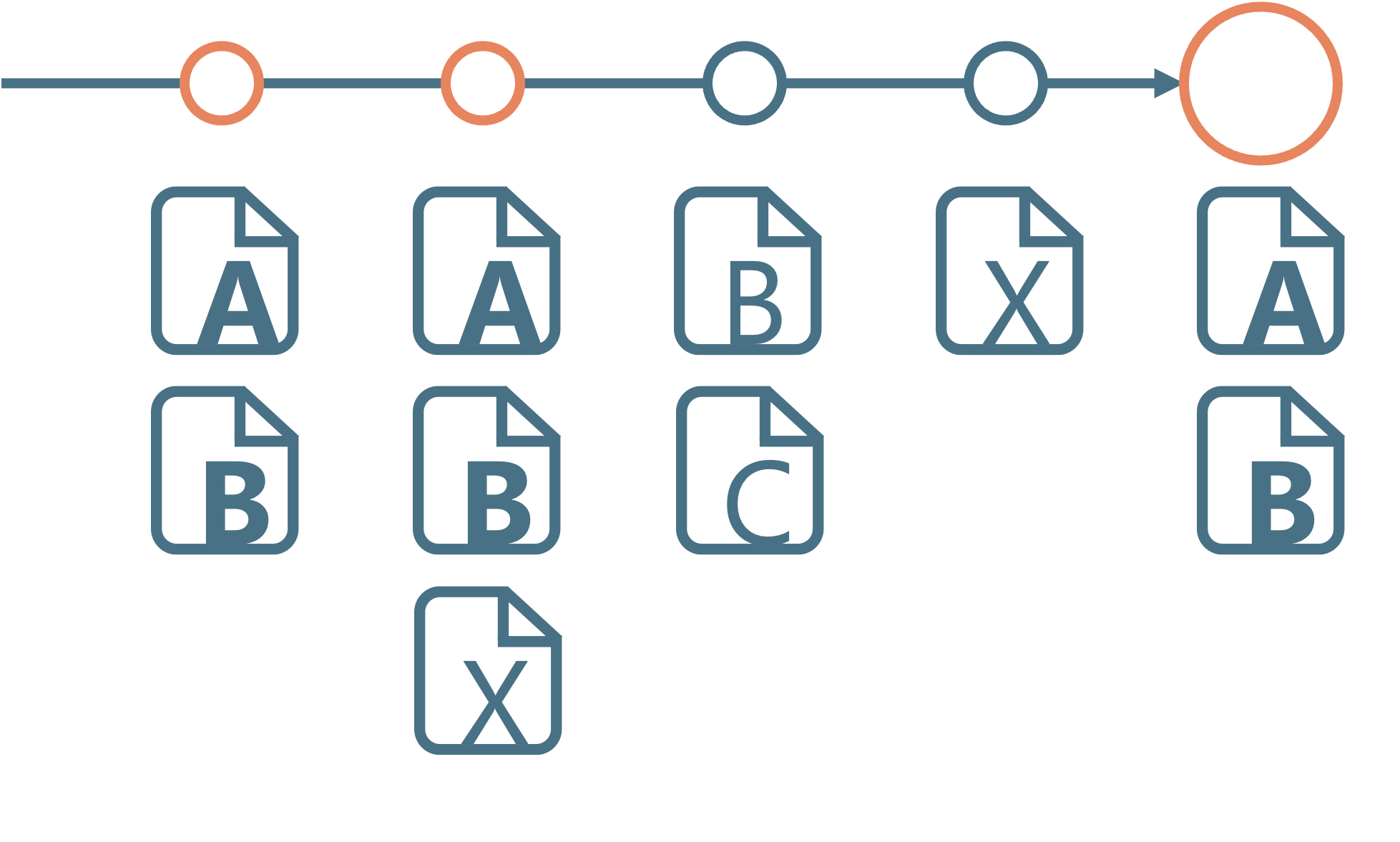}}\hspace{0.25em}
    \subcaptionbox{Data Dependency\label{fig:data_dep}}[.19\linewidth]{\includegraphics[width=.75\linewidth]{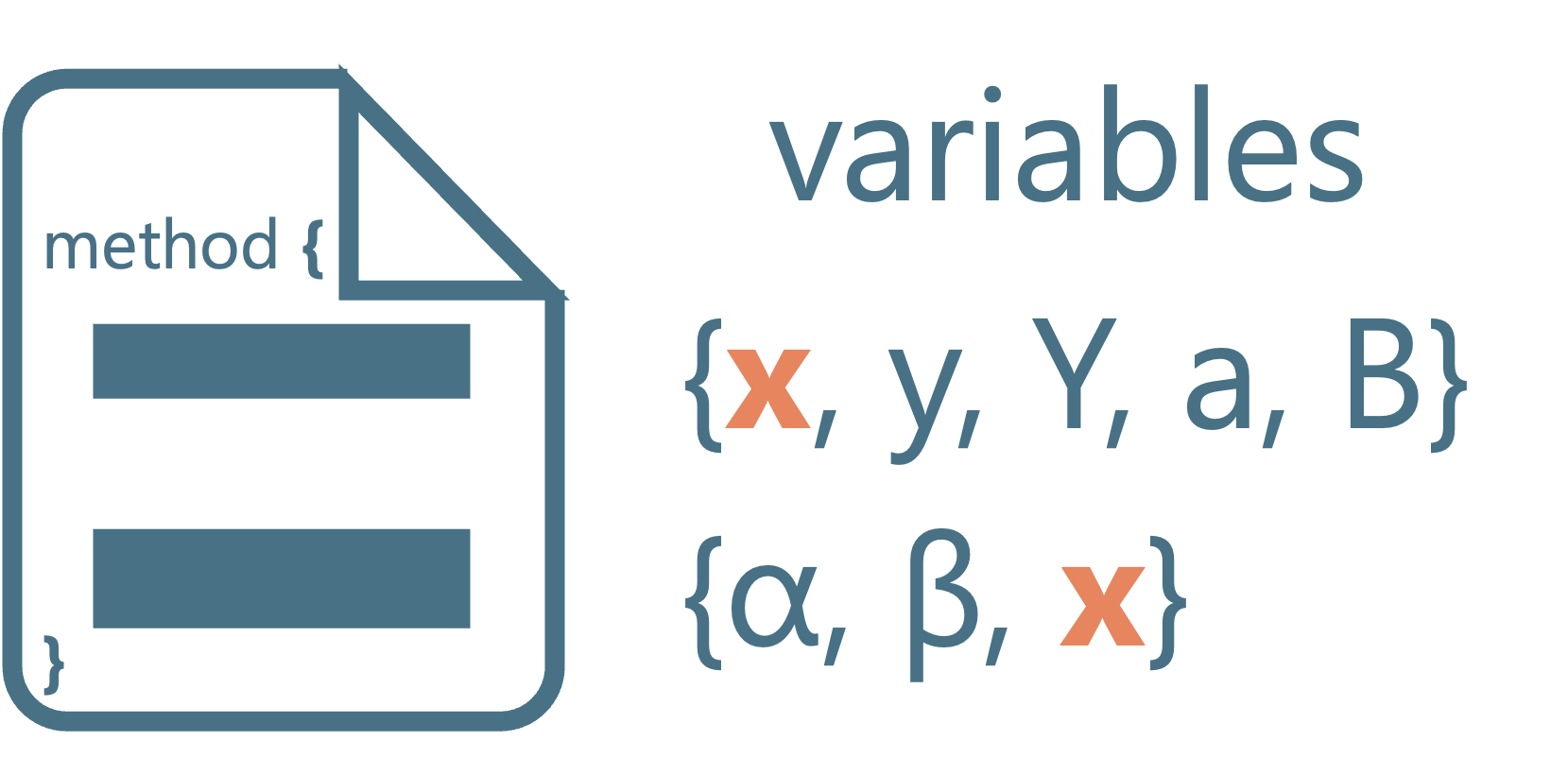}}
    \caption{Conceptual visualizations of the five Confidence Voters.}
    \label{fig:conf_voters}
\end{figure*}

All the six one-tailed KS tests confirm that these observed differences are statistically significant to $p < 0.02$.
Thus, all the null hypotheses, no difference in the data distributions curated from each rule, were rejected.
Regarding each of the Cliff's Deltas, two metrics (the changed LOC, and the minimum File Distance) showed negligible effect size, while the others showed small ones.

These findings suggest that data curated from PRs tends to be more dispersed (our min and max file distance is larger than that of \HerzigRulesPlus).
We do acknowledge that the min file distance Cliff's Delta is negligible; however, the max still suggest more disconnected chunks are present.
We also note that our approach creates CCs that span more projects, further facilitating long range dependencies as a crucial aspect for untangling.
We observe that \HerzigRulesPlus created more clustered chunks, which was also previously examined by Fan \etal~\cite{conf:FanEtal2024ASE_HDGNN} to motivate their hidden dependencies as previous work missed CCs where the STCs were more spread out.
Unlike \HerzigRulesPlus, our rules can also identify CCs confined to a single file, thus CCs consisting of a single file.

Thus, data constructed by the two rules are statistically different along important dimensions suggesting that we should diversify how we collect CC data.
Further, we observe a move to pull-based development on platforms such as GitHub~\cite{article:decan2022use} suggesting that a PR based CC collection is more representative of such projects.
We now show (RQ4) that this impacts Herzig \etal's original approach, hinting that previous work may under-report performance relative to our proposed dataset.

\subsection{{\sf RQ4}: Impact on Automatic Untangling}\label{sec:rq4}

This RQ addresses the most critical aspect of our research: does the data drift we observe in RQ3 impact automatic commit untangling.
Towards this end, we employ the confidence voter based approach from Herzig \etal~\cite{conf:HerzigZeller2013MSR, journal:HerzigEtal2016ESE} and assess the performance on the two datasets.
We focus on the Herzig \etal's approach due to it being applicable without requiring training; learning-based approaches would require retraining and a proper assessment needs the original training data and our new data to properly assess.
While porting Flexeme, which is clustering based, would require extensive static analysis at various project versions.
We leave this to future work.

\subsubsection{Experimental Methodology}
The experiment is designed to isolate the source dataset as the sole variable to directly compare the impact of data drift.
The datasets are the same as in RQ3, and we use the two datasets as test data.
We do not require training data as previously discussed.

We did not use the dataset by Li \etal~\cite{conf:LiEtal2022FSE_UTANGO}, because it was distributed only in graph format, and the original source code was unrecoverable, while the dataset by Fan \etal~\cite{conf:FanEtal2024ASE_HDGNN} which was manually validated would be a small population.

\noindent
\textbf{\emph{Untangling Algorithm:}}
To ensure a fair and direct comparison, we employ the original multi-predictor untangling algorithm proposed by Herzig \etal~\cite{conf:HerzigZeller2013MSR, journal:HerzigEtal2016ESE}.
This algorithm decomposes a change set into a series of individual change operations, which we refer to as chunks.
To be precise, the chunks are not merely arbitrary ones; for the purpose of computing voters such as Call Graph scores, they are partitioned at type (e.g., class or interface) or method boundaries~\cite{conf:LunaEtal2018SOFSEM_diff-region}.
It then uses a set of five scores, known as Confidence Voters, to compute a confidence score between $0$ and $1$ for each pair of chunks, indicating their likelihood of belonging to the same task.

\Cref{fig:conf_voters} illustrates the five Confidence Voters.
For completeness, we now present their detailed definitions:
\begin{itemize}
    \item \textbf{File Distance}:
    This voter measures the proximity of two chunks within the same file.
    If two chunks are in different files or changed types (add or delete), the score is $0$.
    Otherwise, the score is calculated as $1 - (d / l)$, where $d$ is the number of lines between the two chunks, and $l$ is the total number of lines in the file after the changes.
    For example, if two chunks are 10 lines apart in a 100-line file, the score is $1 - (10 / 100) = 0.9$, indicating a high likelihood of being related.
    \item \textbf{Package Distance}:
    This voter measures the semantic distance between the packages of two chunks located in different files.
    If two chunks are in the same file, the score is $0$.
    The score is calculated as $c / m$, where $c$ is the number of common parent package segments from the root, and $m$ is the total number of segments in the longer package path.
    For example, for chunks in ``{\sl com.a.b.c.File1}'' and ``{\sl com.a.b.d.e.File2}'', the common part is ``{\sl com.a}'', so $c=3$.
    The longer path has 5 segments.
    The score is $3 / 5 = 0.60$.
    \item \textbf{Call Graph}:
    This voter analyses the relationships between methods modified by two chunks using a static call graph.
    The distance between two method nodes is defined as the sum of all edge weights along the shortest path between them.
    The weight of an edge between two methods is calculated as the reciprocal of the number of method calls between them.
    A shorter path distance implies a stronger relationship.
    The final score is normalized to a value between 0 and 1 using a sigmoid function, where a shorter distance results in a higher score.
    For instance, if one chunk modifies a method that directly calls another method modified by a second chunk, their path distance will be short, and their score will be high.
    \item \textbf{Change Coupling}:
    This voter leverages the project's history to determine if the files modified by two chunks have frequently been changed together in past commits.
    This concept is based on the work of Zimmermann \etal~\cite{conf:ZimmermannEtal2005OOPSLA_eROSE}.
    The score is calculated as $N_{co} / \max(N_1, N_2)$, where $N_{co}$ is the number of past commits where both files were changed, and $N_1$ and $N_2$ are the total number of commits where each file was changed, respectively.
    A higher score suggests a strong historical coupling.
    In our implementation, this is calculated only for chunks in different files.
    \item \textbf{Data Dependency}:
    This voter checks for direct data dependencies between two chunks within the same method.
    If one chunk writes to a variable that another chunk reads or writes, or if both chunks read the same variable, they are considered dependent, and the score is 1.
    Otherwise, the score is 0.
    Our implementation focuses on dependencies within a single method for simplicity.
    For example, if one chunk changes the initialization of a variable `x` and another chunk uses `x` in a subsequent calculation, their score is 1.
\end{itemize}

After computing the five confidence scores for a given pair of chunks, the algorithm determines their relationship based on a simple thresholding mechanism.
Specifically, it takes the maximum score among the five voters.
If this maximum value is greater than or equal to a threshold (set to 0.5 in this study), the two chunks are considered to belong to the same STC.
If the maximum score is below the threshold, they are assigned to different commits.

By using the same best effort reproduction, we can directly attribute any performance differences to the data drift.

\begin{figure}[t]
    \begin{center}
        \includegraphics[width=0.95\columnwidth]{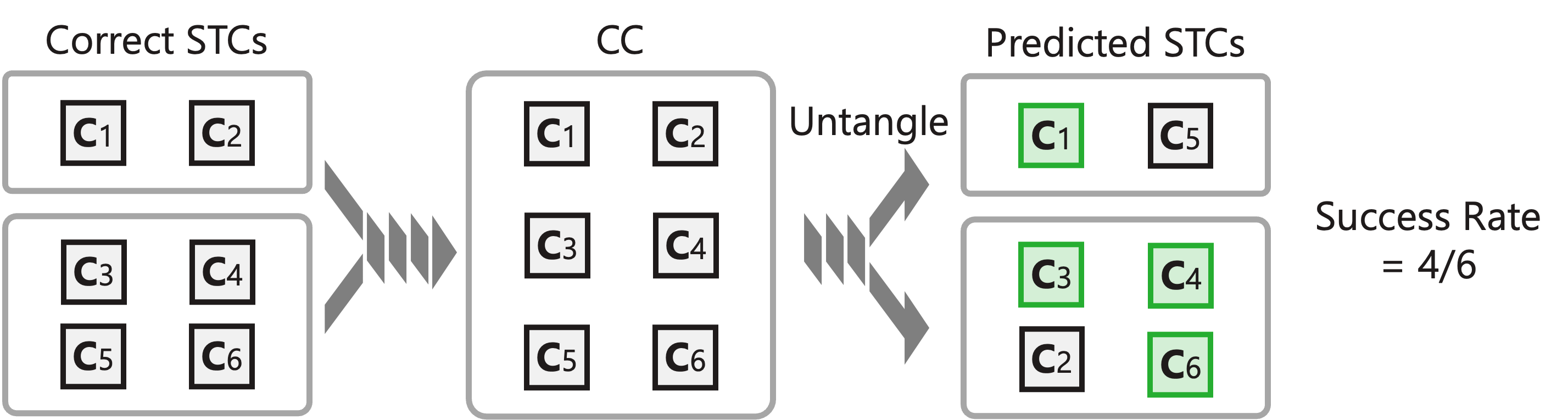}
    \end{center}
    \caption{Example of precision calculation.}
    \label{fig:Ex_SR}
\end{figure}

\noindent
\textbf{\emph{Evaluation Metric:}}
We use the \textit{precision} metric, as defined and used by Herzig \etal~\cite{conf:HerzigZeller2013MSR, journal:HerzigEtal2016ESE}.
This metric is calculated as the proportion of individual chunks that are assigned to the correct partition (i.e., the correct STC), as illustrated in \Cref{fig:Ex_SR}.
A higher precision signifies a more accurate untangling performance.
\begin{equation}
    \textrm{precision} = \frac{\textrm{\# correctly assigned chunks}}{\textrm{Total \# chunks in a commit}}
\end{equation}

\subsubsection{Results and Discussion}
\Cref{table:result_SuccessRate} presents the untangling performance of the algorithm for both conditions.

The results demonstrate that the same algorithm performs significantly differently (Mann-Whitney U, $p < 0.001$) on the two datasets, highlighting that proposed approaches may under-report their performance when applying the Herzig \etal's heuristics to construction datasets.
This finding raises a critical concern for the research community.
Specifically, this data drift threatens the external validity of new approaches that are validated exclusively using data constructed with the original heuristics: while we expect approaches to under-report performance, it remains an unexplored concern.
We expect this to be especially pertinent when developers should apply proposed approaches to pull-based development projects. 
Further, our proposed collection technique can be extended to include newer data, and allows future work to better track in-the-wild CC distributions.

\begin{table}[t]
    \caption{Comparison of untangling performance (precision).}
    \label{table:result_SuccessRate}
    \centering
    \normalsize{
        \begin{tabular}{lccc}
            \hline
            Test Data & Avg. & Mdn. & Std. Dev. \\
            \hline
            Ours & \textbf{0.575} & \textbf{0.571} & \textbf{0.212} \\
            Herzig \etal's & 0.504 & 0.500 & 0.222 \\
            \hline
        \end{tabular}
    }
\end{table}

\section{Implications}\label{sec:discussion}

In this section, we first address implications from each RQ before considering a more holistic perspective.

\subsection{Implcations from our Research Questions}

Our empirical evaluation provides strong, multifaceted support for our automated data curation approach.

\textbf{RQ1 (Dataset Quality)} demonstrated that our filtering method produces high-quality data.
    With a substantial inter-rater agreement ($\kappa=0.76$), human experts confirmed that 76-86\% of our curated PRs are valid CCs and 81-86\% of their commits are valid STCs.
    This result validates our core hypothesis that filtered PRs can serve as reliable proxies.

\textbf{RQ2 (Rule Quality)} proved that our proposed conditions efficiently eliminated data that is not suitable for training, increasing the rate of ideal PRs significantly from 9.5\% to 55.0\%.
    The result from this RQ contributes to preparing supervision CC and STC pair data in Commit Untangling Tasks.
    
\textbf{RQ3 (Dataset Construction Comparison)} showed that a dataset sourced from PRs differed statistically significantly from datasets constructed using the previous Herzig \etal rules along critical metrics. 
    The observed differences were small, however, it suggests that PR-sourced STCs are more dispersed and may cover different methods. 
    Later RQ4 shows that this does have an impact on the confidence-voters-based untangling approach.
    This result suggests that our approach, by using data from modern PR-based workflows, captures a different representation of the structure of CCs in OSS practice.
    Moreover, unlike previous curation rules, our method can identify CCs that encompass a single file.

\textbf{RQ4 (Impact on Automatic Untangling)} showed that the data drift is of practical importance.
    While the result does suggest existing work may under-report performance, this effect remains unexplored by existing work.
    Further, we expect such untangling algorithms to be deployed in pull-based development projects or indeed integrated in CI/CD~\cite{conf:PartachiEtal2020FSE_Flexeme}.

\subsection{Implications of a Scalable Curation Framework}
The most significant contribution of this work is a scalable and adaptable framework for dataset creation, which addresses the long-standing data scarcity problem in this domain.
Our dataset can be expanded by simply extending the data collection horizon (\Cref{sec:approach}, Req. 4), and requires minimal-to-no manual effort.

\subsubsection{Enabling Advanced, Data-Hungry Models}
State-of-the-art, data-hungry models like HD-GNN \cite{conf:FanEtal2024ASE_HDGNN} have demonstrated great potential, but their advancement has been constrained by the lack of large-scale training data.
Our automated curation method offers a direct approach to mitigate this challenge.
\Cref{table:dataset_size_comparison} quantitatively demonstrates this scalability, showing our curated dataset contains over 5.7 times more CCs and 88.3 times more than STCs than the \Herzigdataset.
This abundance of data enables the proper training and evaluation of complex deep learning architectures for completely actual tasks, which was previously infeasible.

\begin{table}[t]
    \caption{Dataset Size Comparison.}
    \label{table:dataset_size_comparison}
    \centering
    \normalsize{
        \begin{tabular}{lrr}
            \toprule
            Dataset & \# CCs & \# STCs \\
            \midrule
            Herzig \etal~\cite{conf:HerzigZeller2013MSR, journal:HerzigEtal2016ESE} & 2,694 & 386 \\
            Our Curated Dataset (Java) & \textbf{15,320} & \textbf{34,091} \\
            Our Curated Dataset (Python) & \textbf{40,303} & \textbf{91,995} \\
            \bottomrule
        \end{tabular}
    }
\end{table}

\subsubsection{Expanding Research Diversity}
The high cost of manual curation has historically limited datasets to a few well-known Java projects.
Because our automated approach is language-agnostic in its core principles (e.g., filtering by commit count, author), it can be readily adapted to other programming languages and a wider variety of modern projects.
This fosters more diverse and comprehensive empirical studies.

\section{Threats to Validity}\label{sec:threats}

\textbf{Internal Validity.}
A primary threat to internal validity is our core assumption that a filtered PR accurately represents a single logical task (a CC).
While RQ1 provides strong evidence to support this, our heuristic-based filtering conditions (informed by previous empirical work~\cite{conf:HerzigZeller2013MSR, journal:HerzigEtal2016ESE}) may still introduce systematic bias by, for example, excluding certain types of valid but unconventional CCs, we partially alleviate this in RQ2.
We conducted the experiment in RQ3, using \HerzigRulesPlus rather than the original ones, however, the rules were added to create the same cut-offs as in our method and remove the bias that may be due to this filtering.
Additionally, our implementation of the Herzig \etal algorithm in RQ4, while based on a careful interpretation of their paper, could have subtle deviations that might affect the results.
This is mitigated by employing the same reproduction for both datasets explored.

\textbf{External Validity.}
Our study is currently limited to Java and Python projects on GitHub.
Development practices can differ across programming languages and platforms (e.g., GitLab).
The generalisability of our findings to other ecosystems requires further validation.

However, the core principles of our filtering conditions (e.g., commit counts, author analysis) are largely language-agnostic, suggesting the framework is adaptable beyond the languages we study in this paper.
Conditions such as the number of commits, the absence of overlapping changes, and metadata-based checks can be applied to any project using a pull-request workflow.
These principles suggest that our curation framework is fundamentally adaptable to other languages like C\# or JavaScript.
The thresholds may need to be recalibrated to accurately capture the typical scope of an STC in another language ecosystem.

Additionally, our repository selection process was based on star counts, which may bias our dataset towards popular, well-maintained open-source projects.
The findings may not be representative of smaller, less popular projects or proprietary, industrial software systems, where development practices can vary considerably.
\section{Conclusion}\label{sec:concl}

The lack of high-quality, large-scale labelled datasets has hampered the advancement of commit untangling research, particularly for data-intensive models.
Foundational benchmarks, such as the one developed by Herzig \etal, have been invaluable to the community, yet their meticulous manual creation process is inherently difficult to scale, previous work opting to use artificially created data without extensive manual validation.

In this paper, we introduced and validated a novel, automated method to address this challenge by curating Composite Commits (CCs) and their constituent Single-Task Commits (STCs) from merged Pull Requests.
We based our intuition on a PR ideally addressing a single high-level task made of individual constituent steps, offering ``natural'' (CC, STCs) pairs.
While we notice some noise in our data, which is fundamentally difficult to remove automatically, we demonstrate that the noise in such datasets is within the tolerance limits of modern machine learning techniques ($<20$\%) (RQ1) and that our filtering significantly improves the collected data (RQ2).
In RQ3, we show that there is a significant albeit small difference between the min/max file and package distances of our dataset and that created using \HerzigRulesPlus.
This suggests that PR data is more dispersed and spans more packages, requiring long-range dependencies to be considered for untangling.
Finally, and most critically, we demonstrate that, despite the small Cliff's delta, there is significant impact on the Herzig \etal confidence voter technique.
This heightens the threat to the external validity of approaches validated on data constructed using the original heuristic approach (RQ4).

Our contributions are threefold: we propose and validate a scalable data collection methodology that has a low noise rate, we show that our method produces datasets that are statistically different from the previous artificial methods, and we demonstrate that this difference has a significant impact on the evaluation of the original confidence voters based on the untangling algorithm.
Crucially, with the move to more pull-based development, we expect our dataset to be more representative.
This opens the door for wider data collection, and future work can focus on expanding the collection and validating the hypothesis on further language ecosystems, such as JavaScript, TypeScript, C\#, or C++.

\subsubsection*{Data Availability}
We make our dataset, intermediate results, and analysis results available at \url{https://figshare.com/s/78ad9367cc58e14d22e1}.

\vspace{0.1cm}
\noindent\textbf{Acknowledgment: }
During the preparation of this work the authors used Google Gemini in order to improve readability and language of the work. After using this tool/service, the authors reviewed and edited the content as needed and take full responsibility for the content of the publication.
This work was partly supported by KAKENHI JP22H03567 and JP26K02888.

\bibliographystyle{IEEEtran-tklab}
\bibliography{ref}

\end{document}